\documentclass[10pt,twocolumn]{article}

\usepackage[utf8]{inputenc}
\usepackage{tabularx}
\usepackage{amsthm}
\usepackage{amssymb}
\usepackage{amsmath}
\usepackage{graphicx}
\usepackage[usenames]{color}
\usepackage{multirow}
\usepackage{url}
\usepackage{epic}
\usepackage{eepic}
\usepackage{stmaryrd}

\newlength{\innerboxwidth}

\newtheorem{definition}{Definition}
\newtheorem{theorem}{Theorem}

\newtheorem{lemma}{Lemma}

\newtheorem{fact}{Fact}

\newcommand{\comment}[1]{}

\newcommand{\Pre}{\mathit{pre}}
\newcommand{\Post}{\mathit{post}}

\newcommand{\AND}{\wedge}

\newcommand{\WP}{{\it wp}}
\newcommand{\DEFS}{\mathit{defs}}
\newcommand{\SHIFT}{\mathit{shift}}
\newcommand{\IF}{\mbox{\sc if}}
\newcommand{\UNSHIFT}{\mathit{unshift}}

\newcommand{\SELECT}{\mbox{\sc select}}

\newcommand{\IL}{\mathit{IL}}

\newcommand{\ghost}[1]{{\it #1}^g}

\newcommand{\MS}{{\bf ms}}

\newcommand{\True}{\mbox{\it tt}}
\newcommand{\False}{\mbox{\it  ff}}

\newcommand{\Val}{\mathit{Val}}

\newcommand{\PC}{\mathit{pc}}

\newcommand{\loca}{\ell}

\newcommand{\Stack}{\mathrm{s}}

\newcommand{\transJVM}{\rightarrow_{\mbox{\scriptsize{JVM}}}}

\newcommand{\tupled}[1]{(#1)}

\newcommand{\bmid}{\mid}

\newcommand{\BinOp}{\circ}
\newcommand{\BinRel}{\mathit{r}} 

\newcommand{\Sem}[1]{\mbox{$\parallel$}{{#1}}\mbox{$\parallel$}}

\newcommand{\Size}[1]{\mbox{$|$}{#1}\mbox{$|$}}
\newcommand{\Local}{\mathrm{l}}

\newcommand{\SRT}{\mathit{SRT}}

\newcommand{\Policy}{\mathcal{P}}
\newcommand{\Contract}{\mathcal{C}}

\newcommand{\Inline}{{\cal I}}
\newcommand{\GI}{\ghost{\Inline}}

\begin{document}

\title{A Proof Carrying Code Framework for Inlined Reference Monitors in Java Bytecode}

\author{Mads Dam, Andreas Lundblad\\Royal Institute of Technology, KTH}

\maketitle

\begin{abstract}
We propose a light-weight approach for certification of monitor inlining 
for sequential Java bytecode using proof-carrying code. The goal is to 
enable the use of monitoring for quality assurance at development time, 
while minimizing the need for post-shipping code rewrites as well as 
changes to the end-host TCB. Standard automaton-based security policies 
express constraints on allowed API call/return sequences. Proofs are
represented as JML-style program annotations. This is adequate in our
case as all proofs generated in our framework are recognized in time
polynomial in the size of the program. Policy adherence is proved
by comparing the transitions of an inlined monitor with those of a 
trusted ``ghost'' monitor represented using JML-style annotations.
At time of receiving a 
program with proof annotations, it is sufficient for the receiver to plug 
in its own trusted ghost monitor and check the resulting verification 
conditions, to verify that inlining has been  performed correctly, of the 
correct policy. We have proved correctness of the approach at the Java 
bytecode level and formalized the proof of soundness in Coq. An
implementation, including an application loader running on a mobile device, 
is available, and we conclude by giving benchmarks for two sample applications.
\end{abstract}

%
%

\section{Introduction}
Program monitoring \cite{Lig06,javamac,javamop} is a well-established technique for software quality assurance, used for a wide range of purposes such as performance monitoring, protocol compliance checking, access control, and general security policy enforcement. The conceptual model is simple: Monitorable events by a client program are intercepted and routed to a decision point where the appropriate action can be taken, depending on policy state such as access control lists, or on application history. This basic setup can be implemented in a huge variety of ways. In this paper our focus is monitor inlining \cite{ErlSch00}. In this approach, monitor functionality is weaved into client code in AOP style, with three main benefits:
\begin{itemize}
\item Extensions to the TCB needed for managing execution of the client, intercepting and routing events, and policy decision and enforcement are to a large extent eliminated.
\item Overhead for marshalling and demarshalling policy information between the various decision and enforcement points in the system is eliminated.
\item Moreover, there is no need to modify and maintain a custom API or Virtual Machine.
\end{itemize}
This, however, presupposes that the user can trust that inlining has been
performed correctly. This is not a problem if the inliner is known to be
correct, and if inlining is performed within the users jurisdiction.
But it could of interest to make inlining available as a quality assurance
tool to third parties (such as developers or operators) as well. In this
paper we examine if proof-carrying code can be used to this effect in the
context of Java and mobile applications, to enable richer, history-dependent
access control than what is allowed by the current, static sandboxing regime.

Our approach is as follows: We assume that J2ME applications
are equipped with {\em contracts} that express the provider commitments on
allowed sequences of API calls performed by the application. Contracts are given as security automata in 
the style of Schneider \cite{Sch00} in a simple contract specification 
language ConSpec \cite{AktNal07}.  The contract is compiled into bytecode and inlined
into the application code as in PoET/PSLang \cite{ErlSchb00}, and a proof
is generated asserting that the inlined program adheres to the contract,
producing in the end a self-certifying code ``bundle'' consisting of the 
application code, the contract, and an embedded proof object.

Upon reception the remote device first determines whe\-ther the received
bundle should be accepted for execution, by comparing the received
contract with the device policy. This test uses a simulation or language 
containment test, and is explored in detail by K. Naliuka et al. \cite{Bielova2009340}. 

The contribution of this paper is the efficient representation, generation, 
and checking of proof objects. The key idea is to compare the 
effects of the inlined, untrusted, monitor with a ``ghost''  monitor
which implements the intended contract. A ghost monitor is 
a virtual monitor which is never actually executed, and which is represented
using program annotations.
Such a ghost monitor is readily available by simply interpreting the 
statements of the ConSpec contract as monitor updates performed before 
and after security relevant method calls. No JVM compilation is required 
at this point, since these updates are present solely for proof 
verification purposes.

The states of the two monitors are compared statically through a 
{\em monitor invariant}, expressing that the state of the embedded 
monitor is in synchrony with that of the ghost monitor. This monitor 
invariant is then inserted as an assertion at each security relevant method call. 
The assertions for the remaining program points could then in principle 
be computed using a weakest pre-condition (WP) calculus. Unfortunately, 
there is no guarantee that such an approach would be feasible. 
However, it turns out that it is sufficient to perform the WP computations 
for the inlined code snippets and not for the client code, under some 
critical assumptions:
\begin{itemize}
\item The inlined code appears as contiguous subsequences of the entire 
instruction sequences in the inlined methods.
\item Control transfers in and out of these contiguous code snippets are allowed 
only when the monitor invariant is guaranteed to hold.
\item The embedded monitor state is represented in such a way that a 
simple syntactic check suffices to determine if some non-inlined 
instruction can have an effect on its value.
\end{itemize}
The last constraint can be handled, in particular, by implementing the 
embedded monitor state as a static member of a final security state class. 
The important consequence is that instructions that do not appear in 
the inlined snippets, and do not include {\tt putstatic} instructions to 
the security state field, may  be annotated with the monitor invariant 
to obtain a fully annotated program. This means, in particular, that a simple 
syntactic check is sufficient to eliminate costly WP checks in almost all 
cases and allows a very open-ended treatment of the JVM instruction set. 

\begin{figure}[ht]
\centering\scriptsize
\ifx\JPicScale\undefined\def\JPicScale{1}\fi
\unitlength \JPicScale mm
\begin{picture}(81.25,70)(0,0)
\put(7.5,62.5){\makebox(0,0)[cc]{Bytecode}}

\put(22.5,62.5){\makebox(0,0)[cc]{Contract}}

\linethickness{0.3mm}
\put(7.5,55){\line(0,1){5}}
\put(7.5,55){\vector(0,-1){0.12}}
\linethickness{0.3mm}
\put(22.5,55){\line(0,1){5}}
\put(22.5,55){\vector(0,-1){0.12}}
\put(15,52.5){\makebox(0,0)[cc]{Inliner}}

\put(15,12.5){\makebox(0,0)[cc]{Proof Generator}}

\linethickness{0.3mm}
\put(1.25,15){\line(1,0){27.5}}
\put(1.25,10){\line(0,1){5}}
\put(28.75,10){\line(0,1){5}}
\put(1.25,10){\line(1,0){27.5}}
\linethickness{0.3mm}
\put(15,45){\line(0,1){5}}
\put(15,45){\vector(0,-1){0.12}}
\put(15,42.5){\makebox(0,0)[cc]{Inlined Classes}}

\linethickness{0.3mm}
\put(15,35){\line(0,1){5}}
\put(15,35){\vector(0,-1){0.12}}
\put(15,32.5){\makebox(0,0)[cc]{Ghost Annotator}}

\linethickness{0.3mm}
\put(15,25){\line(0,1){5}}
\put(15,25){\vector(0,-1){0.12}}
\put(15,22.5){\makebox(0,0)[cc]{Classes + Ghost monitor}}

\linethickness{0.3mm}
\put(15,15){\line(0,1){5}}
\put(15,15){\vector(0,-1){0.12}}
\put(15,22.5){\makebox(0,0)[cc]{}}

\linethickness{0.3mm}
\put(1.25,35){\line(1,0){27.5}}
\put(1.25,30){\line(0,1){5}}
\put(28.75,30){\line(0,1){5}}
\put(1.25,30){\line(1,0){27.5}}
\linethickness{0.3mm}
\put(1.25,55){\line(1,0){27.5}}
\put(1.25,50){\line(0,1){5}}
\put(28.75,50){\line(0,1){5}}
\put(1.25,50){\line(1,0){27.5}}
\put(67.5,52.5){\makebox(0,0)[cc]{Ghost Annotator}}

\put(67.5,12.5){\makebox(0,0)[cc]{VC Checker}}

\linethickness{0.3mm}
\put(53.75,15){\line(1,0){27.5}}
\put(53.75,10){\line(0,1){5}}
\put(81.25,10){\line(0,1){5}}
\put(53.75,10){\line(1,0){27.5}}
\linethickness{0.3mm}
\put(67.5,45){\line(0,1){5}}
\put(67.5,45){\vector(0,-1){0.12}}
\put(67.5,42.5){\makebox(0,0)[cc]{Classes + Ghost monitor}}

\linethickness{0.3mm}
\put(75,35){\line(0,1){5}}
\put(75,35){\vector(0,-1){0.12}}
\put(67.5,32.5){\makebox(0,0)[cc]{VC Generator}}

\linethickness{0.3mm}
\put(67.5,25){\line(0,1){5}}
\put(67.5,25){\vector(0,-1){0.12}}
\put(67.5,22.5){\makebox(0,0)[cc]{Verification Conditions}}

\linethickness{0.3mm}
\put(67.5,15){\line(0,1){5}}
\put(67.5,15){\vector(0,-1){0.12}}
\put(67.5,22.5){\makebox(0,0)[cc]{}}

\linethickness{0.3mm}
\put(53.75,35){\line(1,0){27.5}}
\put(53.75,30){\line(0,1){5}}
\put(81.25,30){\line(0,1){5}}
\put(53.75,30){\line(1,0){27.5}}
\linethickness{0.3mm}
\put(53.75,55){\line(1,0){27.5}}
\put(53.75,50){\line(0,1){5}}
\put(81.25,50){\line(0,1){5}}
\put(53.75,50){\line(1,0){27.5}}
\linethickness{0.3mm}
\put(15,5){\line(0,1){5}}
\put(15,5){\vector(0,-1){0.12}}
\put(15,2.5){\makebox(0,0)[cc]{Adherence Proof}}

\put(15,2.5){\makebox(0,0)[cc]{}}

\linethickness{0.3mm}
\put(67.5,5){\line(0,1){5}}
\put(67.5,5){\vector(0,-1){0.12}}
\put(67.5,2.5){\makebox(0,0)[cc]{Valid/Invalid}}

\put(67.5,2.5){\makebox(0,0)[cc]{}}

\linethickness{0.15mm}
\put(42.5,0){\line(0,1){70}}
\linethickness{0.15mm}
\multiput(30,42.5)(5,0){2}{\line(1,0){2.5}}
\linethickness{0.15mm}
\multiput(37.5,42.5)(0,5){4}{\line(0,1){2.5}}
\linethickness{0.15mm}
\multiput(37.5,60)(5,0){5}{\line(1,0){2.5}}
\linethickness{0.15mm}
\multiput(60,55)(0,3.33){2}{\line(0,1){1.67}}
\put(60,55){\vector(0,-1){0.12}}
\linethickness{0.15mm}
\multiput(30,62.5)(4.74,0){10}{\line(1,0){2.37}}
\linethickness{0.15mm}
\multiput(75,55)(0,5){2}{\line(0,1){2.5}}
\put(75,55){\vector(0,-1){0.12}}
\linethickness{0.15mm}
\multiput(25,2.5)(3.57,0){4}{\line(1,0){1.79}}
\linethickness{0.15mm}
\multiput(37.5,2.5)(0,4.12){9}{\line(0,1){2.06}}
\linethickness{0.15mm}
\multiput(37.5,37.5)(4.09,0){6}{\line(1,0){2.05}}
\linethickness{0.15mm}
\put(60,35){\line(0,1){2.5}}
\put(60,35){\vector(0,-1){0.12}}
\put(15,70){\makebox(0,0)[cc]{Code Producer}}

\put(67.5,70){\makebox(0,0)[cc]{Code Consumer}}

\end{picture}
\caption{\label{fig:framework} The architecture of our PCC implementation.}
\end{figure}
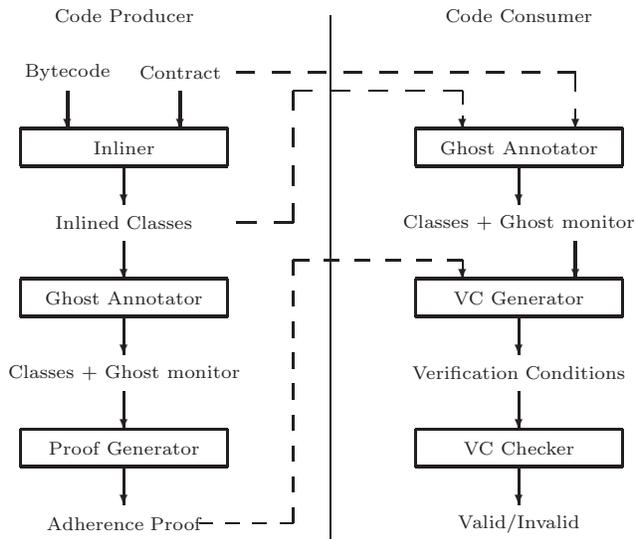

The resulting annotations are 
locally valid in the sense that method pre- and post-conditions match, 
and that each program point annotation follows from successor point 
annotations by elementary reasoning. This allows to robustly and efficiently
generate and check assertions using a standard verification condition (VC) 
approach, as indicated in Figure \ref{fig:framework}.

Our approach is general enough to handle a wide range of inliners. The developer (who has a better insight in the application in question) is free to tweak the inlining process for his specific application and could for instance optimize for speed in certain security relevant call sites, and for code size elsewhere.


\subsection{Related Work}
Our approach adopts the Security-by-Contract (SxC) para\-digm (cf. \cite{Bielova2009340,DraMasNalSia07,Desmet200825,javamac,javamop}) which has been explored and developed mainly within the S$^3$MS project \cite{s3ms}.

Monitor inlining has been considered by a number of authors, cf.~\cite{ErlSch00,ErlSchb00,Erli04,AktDamGur08,VanPie08}.

Erlingsson and Schneider \cite{ErlSchb00} represents security automata directly as Java code snippets, making the resulting code difficult to reason about. The ConSpec contract specification language used here is for tractability restricted to API calls and (normal or exceptional) returns, and uses an independent expression syntax. This corresponds roughly to the call/return fragment of PSLang which includes all policies expressible using Java stack inspection \cite{ErlSch00}.


Edit automata \cite{LigBauWal05,Lig06} are examples of security automata that go beyond pure monitoring, as truncations of the event stream, to allow also event insertions, for instance to recover gracefully from policy violations. This approach has been fully implemented for Java by J. Ligatti et al. in the Polymer tool~\cite{BauLig05} which is closely related to Naccio~\cite{EvaTwy99} and PoET/PSLang~\cite{ErlSchb00}.

Certified reference monitors has been explored by a number of authors, mainly through type systems, e.g. in \cite{SkalkaSmith04,Bauer02typesand,Wal00,HamMor06,RobMan01}, but more recently also through model checking and abstract interpretation \cite{Srid10,Sridb10}. Directly related to the work reported here is the type-based Mobile system due to Hamlen et al. \cite{HamMor06}. The Mobile system uses a simple library extension to Java bytecode to help managing updates to the security state. The use of linear types allows a type system to localize security-relevant actions to objects that have been suitably unpacked, and the type system can then use this property to check for policy compliance. Mobile enforces per-object policies, whereas the policies enforced in our work (as in most work on IRM enforcement) are per session. Since Mobile leaves security state tests and updates as primitives, it is quite likely that Mobile could be adapted, at least to some forms of per session policies. On the other hand, to handle per-object policies our approach would need to be extended to track object references. Finally, it is worth noting that Mobile relies on a specific inlining strategy, whereas our approach, as mentioned in the previous section, is less sensitive to this.

In \cite{Srid10,Sridb10} Sridhar et al. explores the idea of certifying inlined reference monitors for ActionScript using model-checking and abstract interpretations. The approach is not tied to a specific inlining strategy and is general enough to handle different inlining techniques including non-trivial optimizations of inlined code. Although the certification process is efficient, the analysis however, has to be carried out by the consumer.

For background on proof-carrying code we refer to \cite{Nec97}. Our approach is based on simple Floyd-like program point annotations in the style of Bannwarth and M\"uller \cite{BanMue05}, and method specifications extended by pre- and post-conditions in the style of JML \cite{jmldbc}. Recent work related to proof-carrying code for the JVM include \cite{mobiuspcc}, all of which has been developed in the scope of the Mobius project.

An alternative to inline reference monitoring and proof-carrying code, is to produce binaries that are structurally simple enough for the consumer to analyze himself. This is currently explored by B. Chen et al. in the Native Client project~\cite{nacl} which handles untrusted x86 native code. This is done through a customized compile chain that targets a subset of the x86 instruction set, which in effect puts the application in a sandbox. When applicable it has a few advantages in terms of runtime overhead, as it eliminates the monitoring altogether, but is constrained in terms of application and policy complexity.


\subsection{Overview of the Paper}
The JVM machine model is presented in Section 2. In Section 3 the state assertion language is introduced, and in Section 4 we address method and program annotations and give the conditions for (local and global) validity used in the paper. We briefly describe the ConSpec language and (our version of) security automaton in Section 5. The example inlining algorithm is described briefly is Section 6. Section 7 introduces the ghost monitor, and Section 8, then, presents the main results of the paper, namely the algorithms for proof generation and proof recognition, including soundness proofs. Finally, Section 9 reports briefly on our prototype implementation, and we conclude by discussing some open issues and directions for future work.

\section{Program Model}\label{sect:prog_model}
We assume that the reader is familiar with Java bytecode syntax and the Java Virtual Machine (JVM). Here, we only present components of the JVM, that are essential for the definitions in the rest of the text. Much of this is standard and can be skipped in a first reading. A few simplifications have been made in the presentation. In particular we disregard static initializers, and to ease notation a little we ignore issues concerning overloading. We use $c$ for (fully qualified) class names, $m$ for method names, and $f$ for field names. Types are either primitive or object types, i.e. classes, or arrays. Class declarations induce a class hierarchy, denoted by $<:$. If $c$ defines $m$ (declares $f$) explicitly, then $c$ defines (declares) $c.m$ ($c.f$). Otherwise, $c$ defines $c'.m$ (declares $c'.f$) if $c$ is the smallest superclass of $c'$ that contains an explicit definition (declaration) of  $c.m$ ($c.f$). Single inheritance ensures that definitions/declarations are unique, if they exist.

We let $v$ range over the set of all values of all types. Values of object type are (typed) locations, mapped to objects, or arrays, by a heap $h$. The typing assertion $h\vdash v : c$ asserts that $v$ is some location $\loca$, and that in the typed heap $h$, $\loca$ is defined and of type $c$, and similarly for arrays. Typing preserves the subclass relation, in the sense that if $h\vdash v:c$ and $ c <: c'$ then $h\vdash v:c'$ as well. For objects, it suffices to assume that if $h\vdash v:c$ then the object $h(v)$ determines a field $h(v).f$ (method $h(v).m$) whenever $f$ ($m$) is declared (defined) in $c$. Static fields are identified with field references of the form $c.f$. To handle those, heaps are extended to assignments of values to field references.

A program is a set of classes, and for our purposes each class denotes a mapping from method identifiers to definitions $(I,H)$ consisting of an instruction array $I$ and an exception handler array $H$.

We write $c.m[L] = \iota$ to indicate that $c(m)=(I,H)$ and that $I_L$ is defined and equal to the instruction $\iota$. The exception handler array $H$ is a list of of exception handlers. An exception handler $\tupled{b, e, L, c}$ catches exceptions of type $c$ and its subtypes raised by instructions in the range $[b, e)$ and transfers control to address $L$, if it is the first handler in the handler array that catches the exception for the given type and instruction.

A \emph{configuration} of the JVM is a pair $C= (h, R)$ of a heap $h$ and a stack $R$ of activation records. For normal execution, the activation record at the top of the execution stack has the shape $(M,\PC, s, l)$, where $M$ is the currently executing method, $\PC$ is the program counter, $s\in\Val^*$ is the operand stack, and $l$ is the local variable store. Except for API calls (see below) the transition relation $\transJVM$ on JVM configurations is standard. A configuration $(h, (M,\PC,s,l)::R)$ is {\em calling}, if $M[\PC]$ is an invoke instruction, and it is {\em returning normally}, if $M[\PC]$ is a return instruction. For exceptional configurations the top frame has the form $(\ell)$ where $\ell$ is the location of an exceptional object, i.e. of class Throwable. Such a configuration is called {\em exceptional}. We say that $C$ is {\em returning exceptionally} if $C$ is exceptional, and if $C\transJVM C'$ implies that $C'$ is exceptional as well. I.e. the normal frame immediately succeeding the top exceptional frame in $C$ is popped in $C'$, if $C'$ is exceptional as well.

An \emph{execution} $E$ of a program $P$ is a (possibly infinite) sequence of JVM configurations $C_0 C_1 \ldots$ where $C_0$ is an initial configuration consisting of a single, normal activation record with an empty stack, no local variables, $M$ as a reference to the main method of $P$, $\PC=0$ and for each $i\geq 0$, $C_i \transJVM C_{i+1}$. We restrict attention to configurations that are \emph{type safe}, in the sense that heap contents match the types of corresponding locations, and that arguments and return/exceptional values for primitive operations as well as method invocations match their prescribed types. The Java bytecode verifier serves, among other things, to ensure that type safety is preserved under machine transitions.

The only non-standard aspect of $\transJVM$ is the treatment of API methods. We assume a fixed API for which we have access only to the signature (types), but not the implementation, of its methods. We therefore treat API method calls as atomic instructions with a non-deterministic semantics. In this sense, we do not practice \emph{complete mediation}~\cite{saltzer75}. When an API method is called either the $\PC$ is incremented and arguments popped from the operand stack and replaced by an arbitrary return value of appropriate type, or else an arbitrary exceptional activation record is returned. Similarly, 
the return configurations for API method invocations contain an arbitrary heap, since we do not know how API method bodies change heap contents. Our approach hinges on our ability to recognize API calls. This property is destroyed by the \emph{reflect} API, which is consequently not considered.

\section{Assertions}
Annotations are given in a language similar to the one described by F. Y. Bannwart and P. M{\"u}ller in \cite{BanMue05}. 
The syntax of assertions $a$ and (partial) expressions $e$ 
are given in the following BNF grammar:
\[
\begin{array}{l@{~::=~}l}
e & v \bmid e.f \bmid c.f \bmid \Stack_i \bmid \Local_i \bmid e \BinOp e \bmid a \rightarrow e | e \bmid(e,e) \bmid \bot \\[2mm]
a & \True \bmid \False \bmid e\ \BinRel\ e \bmid a \AND a \bmid \neg a \bmid e : c
\end{array} \]
where $i\in\omega$. The semantics, as mappings $\Sem{e}C$ and $\Sem{a}C$ is given in Figure \ref{semantics}.
\begin{figure}
$\begin{array}{r@{~=~}l@{}}
\Sem{e.f}(h, R) & h(\Sem{e}(h, R)).f \\
\Sem{c.f}(h, R) & h(c.f) \\
\Sem{\Stack_i}(h, (M,\PC,s,r)::R) & s_i \\
\Sem{\Local_i}(h, (M,\PC,s,r)::R) & l_i \\
\Sem{e_1 \BinOp e_2}C & \Sem{e_1}C \BinOp \Sem{e_2}C \\
\Sem{e_1 \rightarrow e_2 \mid e_3}C & \left\{\begin{array}{@{}ll}
                                      \Sem{e_2}C\ , & \Sem{e_1}C = \True \\
                                      \Sem{e_3}C, & \mbox{otherwise}
                                      \end{array}\right. \\
\Sem{(e_1,e_2)}C & (\Sem{e_1}C,\Sem{e_2}C) \\
\Sem{\bot}C & \bot \\
\Sem{\True} & \True \\
\Sem{\False} & \False \\
\Sem{e_1\ \BinRel\ e_2}C & \Sem{e_1}C\ \BinRel\ \Sem{e_2}C \\
\Sem{a_1\wedge a_2}C & \Sem{a_1}C \wedge \Sem{a_2}C \\
\Sem{\neg a}C & \overline{\Sem{a}C} \\
\Sem{e : c}(h, R) & \left\{\begin{array}{@{}ll} \True & 
                   \mbox{if } h\vdash \Sem{e}(h, R) : c \\
                   \False & \mbox{otherwise} \end{array}\right. \\
\end{array}$
\caption{\label{semantics}Semantics of expressions and assertions}

\end{figure}
The operations $\BinOp$ and $\BinRel$ are generic binary operators/relation symbols, respectively, with Kleene equality. The expression $\Stack_i$ refers to the $i$'th element of the operand stack, and $\Local_i$ refers to the $i$'th local variable; the expression $a \rightarrow e_1 \mid e_2$ is a conditional, $(e_1,e_2)$ is pairing and $\True$ and $\False$ represent true and false respectively; a {\em heap assertion} is an assertion that does not reference the stack, or any of the local variables. Disjunction ($\vee$) and implication ($\Rightarrow$) are defined as usual. We let $ \IF(a_0, a_1, a_2) $ denote the conditional expression $ (a_0 \Rightarrow a_1) \AND (\neg a_0 \Rightarrow a_2) $ and $\SELECT(\mathbf{a_1}, \mathbf{a_2}, a_{\it else})$ the generalized conditional expression $\IF(a_{1,0}, a_{2,0}, \IF(a_{1,1}, a_{2,1}, \ldots, \IF(a_{1,n}, a_{2,n}, a_{\it else})\ldots))$.

\section{Extended Method Definitions}
In this section we extend the method definitions by an array of program point assertions and by invariants at method entry and (normal or exceptional) return.
\begin{definition}[Extended Method Definition]
An {\em extended method definition} is a tuple $(I, H, A, \Pre,\Post)$ in which $(I,H)$ is a method definition, $A$ is an array of assertions such that $\Size{I}=\Size{A}$ and $\Pre$ and $\Post$ are heap assertions. An {\em extended program} is a program with extended methods.
\end{definition}
For extended programs, the notions of transition and 
execution are not affected by the presence of assertions. An extended 
program is valid, if all annotations are validated by their corresponding configurations in any execution starting in a configuration satisfying the initial pre-condition. In other words:
\begin{definition}[Extended Program Validity]
An extended program $P$ is {\em valid} if for each maximal execution $E=C_0 C_1 \cdots C_k$ of $P$
\begin{enumerate}
\item $\Sem{\Pre_{\mbox{\scriptsize\tt main}}}C_0$ holds,
\item $\Sem{\Post_{\mbox{\scriptsize\tt main}}}C_k$ holds, and
\item for each $i$ such that $0\leq i \leq k$, if $C_i$ has the shape $((M,\PC,s,r)::R,h)$ and $P(M) = (I,H,A,\Pre,\Post)$ then $\Sem{A_\PC}C_i$ holds
\end{enumerate}
\end{definition}
The WP-calculus used in the proof generation / recognition is given in Table \ref{tab:wp}.
%
%
The definition uses the auxillary functions $\SHIFT$ and $\UNSHIFT$ which 
increments, resp. decrements, each stack index by one and $\DEFS(c.m)$ which 
denotes the set of all classes $c'$ such that $c <: c'$ and $c'$ defines $m$.
\begin{table}
$\begin{array}{l@{~}l@{}}
\hline
I_L                        & \WP_{(I, H, A, \Pre, \Post)}(L) \\ \hline
\mbox{\tt instanceof}\ c   & A_{L+1}[\Stack_0 : c/\Stack_0] \\
\mbox{\tt aload}\ n        &  \UNSHIFT(A_{L+1}[\Local_n/\Stack_0]) \\
\mbox{\tt astore}\ n       & (\SHIFT(A_{L+1}))\wedge \Stack_0 = \Local_n \\
\mbox{\tt athrow}          & \mbox{\small\SELECT}((\Stack_0:c\wedge b\leq L < e)_{(b,e,L',c)\in H} \\
                           & \ \ \ (A_{L'})_{(b,e,L',c)\in H},\Post) \\
\mbox{\tt dup}             & \UNSHIFT (A_{L+1}[\Stack_1/\Stack_0]) \\
\mbox{\tt getfield}\ f     & \UNSHIFT (A_{L+1}[\Stack_0.f/\Stack_0]) \\
\mbox{\tt getstatic}\ c.f  & \UNSHIFT (A_{L+1}[c.f/\Stack_0]) \\
\mbox{\tt goto}\ L'        & A_{L'}  \\
\mbox{\tt iconst\_}n       & \UNSHIFT(A_{L+1}[n/\Stack_0]) \\
\mbox{\tt if\_icmpeq}~L'   & \IF(\Stack_0=\Stack_1,\SHIFT^2(A_{L'}),\SHIFT^2(A_{L+1}))\\
\mbox{\tt ifeq}\ L'        & \IF(\Stack_0=0,\SHIFT(A_{L'}), \SHIFT(A_{L+1})) \\
\mbox{\tt invoke}\ c.m     & \bigwedge_{c' \in \mathit{defs}(c.m)} \Pre_{c'.m} \\
\mbox{\tt putstatic}\ c.f  & \SHIFT(A_{L+1})[\Stack_0/c.f] \\
\mbox{\tt return}          & \Post \\
\mbox{\tt ldc}\ v          & \UNSHIFT(A_{L+1}[v/\Stack_0]) \\
\mbox{\tt invokestatic}    & \\
\ \ \ \ \mbox{\tt System.exit} & \True
\end{array}$
\caption{Specification of the $\WP_M$ function}
\label{tab:wp}
\end{table}
The account of dynamic call resolution in Table \ref{tab:wp} is crude, 
but the details are unimportant since, in this paper, pre- and 
post-conditions are always identical and common to all methods.

A locally valid method is one for which each assertion can be validated
by reference to ``neighbouring'' assertions only.
%
%
%
\begin{definition}[Local Validity]\label{def:local_validity}
An extended method $M = (I, H, A, \Pre, \Post)$ is locally valid, if
the verification conditions
\begin{enumerate}
  \item $\Pre \Rightarrow A_0$, and
  \item $A_L \Rightarrow \WP_M(L)$ for all $0 \leq L < |I|$
\end{enumerate}
are valid. 
An extended program is locally valid if all its methods are locally valid 
and the pre-condition of the main method holds in an initial configuration.
\end{definition}
We note that local validity implies validity, as expected.
\begin{theorem}[Local Validity Implies Validity]\label{thm:loc_impl_glob}
For any extended program $P$, if $P$ is locally valid then $P$ is valid.
\end{theorem}
\begin{proof}
Follows by induction on the length of the execution. For details we refer to the Coq formalization \cite{coqscript}.
\end{proof}

\section{Security Specifications}\label{sect:policies}
We consider security specifications written in a policy specification 
language ConSpec~\cite{AktNal07}, similar to 
PSlang~\cite{ErlSchb00}, but more constrained, to be amenable to 
analysis. An example specification is given in Figure 
\ref{fig:conspec_example}. The syntax is intended to be largely 
self-explanatory: The specification in Figure \ref{fig:conspec_example} 
states that the program can only send files using the Bluetooth Obex 
protocol upon direct request by the user. No exception may arise 
during evaluation of the user query.
\begin{figure}
\centering
{\ttfamily
\begin{tabular}{@{}l@{}}
SECURITY STATE String lastApproved = "";\\
\\
AFTER file = GUI.fileSendQuery()\\
\phantom{~~~~}PERFORM true -> \verb+{+ lastApproved = file; \verb+}+\\
\\
EXCEPTIONAL GUI.fileSendQuery()\\
\phantom{~~~~}PERFORM\\
\\
BEFORE Bluetooth.obexSend(String file)\\
\phantom{~~~~}PERFORM file = lastApproved -> \verb+{ }+\\[2mm]
\end{tabular}
}
\caption{\label{fig:conspec_example} A security specification example written in ConSpec.}
\end{figure}

A ConSpec specification tells when and with what arguments an API method 
may be invoked. If the specification has one or more constraints on a 
method, the method is \emph{security relevant}. In 
the example there are two security relevant methods, 
{\tt GUI\!.fileSendQuery} and {\tt Bluetooth\!.\!obexSend}. The specification 
expresses constraints in terms 
{\sc before}, {\sc after} and {\sc exceptional} clauses. Each clause is 
a guarded command where the guards are side-effect free and terminating 
boolean expressions, and the assignment updates the security state. 
Guards may involve constants, method call parameters, object fields, 
and values returned by accessor or test methods that are guaranteed to 
be side-effect free and terminating. Guards are evaluated top to bottom 
in order to obtain a deterministic semantics. If no clause guard holds, 
the policy is violated. In return clauses the guards must be 
exhaustive.

\subsection{Security Automata}
A ConSpec contract determine a security automaton $(Q,\Sigma,$ $\delta, q_0)$ where $Q$ is a countable (not necessarily finite) 
set of states, $\Sigma$ is the alphabet of security relevant actions, 
$q_0\in Q$ is the initial state, and $\delta:Q\times \Sigma \rightarrow Q$ 
is the transition function. We assume a special error state $\bot\in Q$
and view all states in $Q$ except $\bot$ as accepting. We require that
security automata are strict in the sense that $\delta(\bot,\alpha)=\bot$.

A security automata is generated by a ConSpec contract in a straightforward manner (cf.~\cite{AktDamGur08}). The alphabet $\Sigma$ is partitioned into 
pre-actions (for calls) and (normal or exceptional) post-actions 
(for normal or exceptional returns). Pre-actions have the form 
$(c.m, \mathbf{v})^\uparrow$, normal post-actions have the form 
$(c.m, \mathbf{v}, r)^\downarrow$ and exceptional post-actions have 
the form $(c.m, \mathbf{v})^\Downarrow$, where $c.m$ is the 
relevant API-method, $\mathbf{v}$ is the arguments used when 
calling the method, and $r$ is the returned value.

Executions produce security relevant actions in the expected manner.
A calling configuration generates a pre-action determined 
by the called method and the current arguments (top $n$ operand stack 
values for an $n$-ary method). A returning configuration then gives 
rise to a normal post-action determined by the identifier of the 
returning method and the return value (top operand stack value). 
For sake of simplicitly we assume that all API methods return a value. 
An exceptionally returning configuration generates an exceptional 
post-action determined by the method identifier of the returning method. 
The security relevant actions (the security relevant trace) of an 
execution $E$ is denoted by $\SRT(E)$ and formally defined below.

\begin{definition}[Security Relevant Trace] The security relevant trace, 
$\SRT(E)$, of an execution $E$ is defined as
\[
\begin{array}{@{}r@{\ =\ }l@{}}
\SRT(E) & \SRT(E, \epsilon) \\
\SRT(\epsilon, \epsilon) & \epsilon \\
\SRT(CE, \gamma) & \left\{
    \begin{array}{@{}l@{}}
    (c'.m', \mathbf{v})^\uparrow \SRT(E, \mathbf{v}::\gamma) \\[1mm]
    \hspace{4mm} \parbox{5cm}{if $C = (h, (c.m,\PC,\mathbf{v}::s,l)::R)$ is calling $c'.m'$} \\[4mm]
    
    (c.m, \mathbf{v}, r)^\downarrow \SRT(E, \gamma')\\[1mm]
    \hspace{4mm} \parbox{5cm}{if $C = (h, (c.m, \PC, r::s, l)::R)$ is returning  and $\gamma = \mathbf{v}::\gamma'$}\\[7mm]
    
    (c.m, \mathbf{v})^\Downarrow\SRT(E, \gamma') \\[1mm]
    \hspace{4mm} \parbox{5cm}{if $C = (h, (o)::R)$ is returning exceptionally and $\gamma = \mathbf{v}::\gamma'$}\\[7mm]
    
    \SRT(E, \gamma) \qquad \mbox{otherwise}
\end{array}\right.\end{array}\]
\end{definition}

We generally identify a ConSpec contract with its set of security relevant
traces, i.e. the language recognized by its corresponding security automaton.
A program is said to \emph{adhere} to a contract if all its security 
relevant traces are accepted by the contract.
\begin{definition}[Contract Adherence]
The program $P$ adheres to contract $\Contract$ if for all 
executions $E$ of $P$, $\SRT(E)\in\Contract$.
\end{definition}

\section{Example Inliner}\label{sect:inlining}
In this section we give an algorithm for monitor inlining (from now on referred to as an inlining algorithm, or simply an inliner) in the style of Erlingsson~\cite{ErlSch00}. As previously mentioned, the developer is free to decide what inlining strategy to use, so the algorithm presented here serves merely as an example and does for instance not include any optimizations. For the implementation details and an example, we refer to Appendix \ref{app:example_inliner}.

The inliner traverses the instructions and replaces each invoke instruction with a block of monitoring code. This block of code first stores the method arguments in local variables for use in post-actions. Then the class hierarchy is traversed bottom up for virtual call resolution, and when a match is found the relevant clauses, guards, and updates are enacted. For post-actions the main difference is in exception handling; exceptions are rerouted for clause evaluation, and then rethrown.

We refer to the method resulting from inlining a method $M$ (program $P$) with a contract $\Contract$ as $\Inline(M,\Contract)$ ($\Inline(P,\Contract)$). The main correctness property we are after for inlined code is contract compliance:
\begin{theorem}[Inliner Correctness]
The inlined program $\Inline(P,$ ${\Contract})$ adheres to $\Contract$.
\end{theorem}
\begin{proof}
This follows from the fact that we are always able to generate a valid adherence proof (theorem \ref{thm:proof_generation}) and that the existence of such adherence proof ensures contact adherence (theorem \ref{thm:proof_soundness}). (Both statements are proved in later sections.)
\end{proof}

\section{The Ghost Monitor}
The purpose of the ghost monitor is to keep track of what the embedded
monitor state should be at key points during method execution. This provides
a useful reference for verification. Moreover, since the ghost monitor assigns only to 
special ghost variables that are invisible to the client program, and
since it is incapable of blocking, it does not in fact
have any observable effect on the client program.

The ghost monitor uses
special assignments which we refer to as ghost updates: Guarded multi-assignment 
commands used for updating the state of the ghost monitor and for 
storing method call 
arguments and dynamic class identities in temporary variables.
A ghost update has the 
shape $\langle {\bf x}^g := e\rangle $ where ${\bf x}^g$ is a tuple of 
ghost variables, special variables used only by the ghost monitor, and 
$e$ is an expression of matching type. Typically, 
$e$ is a conditional of similar shape as the policy expressions, and 
$e$ may mention security state ghost variables as well as other ghost 
variables holding security relevant call parameters. Given the 
post-condition $A_{L+1}$, the weakest pre-condition for the ghost 
instruction $\langle \ghost{{\bf x}} := e\rangle $ at label $L$ is 
$\WP_M(L)=A_{L+1}[e/\ghost{{\bf x}}]$.

The ghost updates are embedded right before and after each security 
relevant invoke instruction as well as in an exception handler catching 
any exception ({\tt Throwable}) thrown by the invoke instruction and 
nothing else. Note that the existence of such an exception handler is 
easily checked, and that the code delivered 
by our inliner always has exception handlers of this form. The details are 
presented in Figure \ref{fig:ghost_embedding}. A method $M$ with ghost 
updates embedded, corresponding to the security automaton of a 
contract $\Contract$ is denoted by $\GI(M, \Contract)$.
\begin{figure}
\centering
\begin{tabular}{@{}r@{~}l@{}}
$L$: & $\langle (t^g, \mathit{args}^g_1, \ldots, \mathit{args}^g_n) := (s_n, \ldots, s_0)\rangle $ \\
   & \\[-3mm]
   & $\begin{array}[t]{@{}l@{~}c@{~}l@{~}l}
         \langle \MS^g &:=& t^g : c^k & \rightarrow \delta(\MS^g, (c^k.m, \mathit{args}^g)^\uparrow) \\[-1.5mm]
                       &  &           & \vdots \\
                       &| & t^g : c^1 & \rightarrow \delta(\MS^g, (c^1.m, \mathit{args}^g)^\uparrow) \\
                       &| & \MS^g\rangle & \\
         \end{array}$ \\
   & \\[-3mm]
   & {\tt invokevirtual~c.m} \\
   & \\[-3mm]
   & $\begin{array}[t]{@{}l@{~}c@{~}l@{~}l}
         \langle \MS^g &:=& t^g : c^k & \rightarrow \delta(\MS^g, (c^k.m, \mathit{args}^g, s_0)^\downarrow) \\[-1.5mm]
                       &  &           & \vdots \\
                       &| & t^g : c^1 & \rightarrow \delta(\MS^g, (c^1.m, \mathit{args}^g, s_0)^\downarrow) \\
                       &| & \MS^g \rangle & \\
         \end{array}$\\ 
   &\\[-3mm]
   &$\vdots$ \\
   &\\[-3mm]
 $L_{\it HStart}$:&$\begin{array}[t]{@{}l@{~}c@{~}l@{~}l}
         \langle \MS^g &:=& t^g : c^k & \rightarrow \delta(\MS^g, (c^k.m, \mathit{args}^g)^\Downarrow) \\[-1.5mm]
                       &  &           & \vdots                                   \\
                       &| & t^g : c^1 & \rightarrow \delta(\MS^g, (c^1.m, \mathit{args}^g)^\Downarrow) \\
                       &| & \MS^g \rangle &
         \end{array}$
\end{tabular}
\caption{\label{fig:ghost_embedding} Ghost updates induced by security automaton $(Q, \Sigma, \delta, q_0)$ for an invokation of $c.m$, where $t^g$ is the target object, $\mathit{args}^g$ represents the arguments and $c^1 <: \ldots <: c^k$ denote all API-classes defining or overriding $m$.}
\end{figure}

Let $\GI(M, \Contract)$ be the result of embedding a ghost monitor
corresponding to contract $\Contract$ into $M$.
The key property of the ghost monitor is that the trace of ghost monitor 
states in an execution $E$, is the same as
the states visited by the security  automaton, given $\SRT(E)$ as input. 
This is easily be shown by an induction over the length of $E$.
\begin{lemma}\label{lem:ghost_mon_correct}
Let $E = C_0\ldots C_k$ be an execution of ${\cal I}^g(P, \Contract)$ 
and $\MS^g_i$ denote the ghost monitor state in 
configuration $C_i$. If for all $0 \leq i \leq k$, $\MS^g_i \neq \bot$, 
then $\SRT(E) \in \Contract$.
\end{lemma}
\begin{proof}
Follows by induction on the length of the execution. For details we refer to the Coq formalization \cite{coqscript}.
\end{proof}

\section{Contract Adherence Proofs}
The key idea of a contract adherence proof is to show that the embedded 
monitor state $\MS$ of the program
${\cal I}^g(P, \Contract)$ and the ghost monitor state $\MS^g$ 
are in agreement at certain program points. 
These points certainly need to 
include all potentially security relevant call and return
sites. But, since we aim for a procedural analysis, and to cater for
virtual call resolution actually all call 
and return sites are included. 

In fact, this is all that is needed, and hence:
\begin{definition}[Adherence Proof]\label{def:proof_validity}
An adherence proof for program $P$ and contract $\Contract$
assigns to each method $M = (I,H)$ in ${\cal I}^g(P, \Contract)$ 
an assertion array $A$ such that the extended method
$(I,H,A,\MS = \MS^g,\MS = \MS^g)$ is locally valid.
\end{definition}
\noindent Such an account has two main benefits which are heavily exploited below:
\begin{itemize}
\item It leaves the choice of a particular proof generation 
strategy open.
\item It opens for a lightweight approach to on-device proof checking, by
performing the local validity check on a program with a locally produced
ghost monitor.
\end{itemize}
\begin{theorem}[Adherence Proof Soundness]\label{thm:proof_soundness}
If an adherence proof exists for a program $P$ and contract $\Contract$, 
then $P$ adheres to $\Contract$.
\end{theorem}
\begin{proof}
Assume $\Pi$ is an adherence proof for a program $P$ and a contract 
$\Contract$. By theorem \ref{thm:loc_impl_glob} we know that the
corresponding extended program for ${\cal I}^g(P, \Contract)$ is
globally valid.
This implies that $\MS = \MS^g$ at each configuration that is calling 
(or returning from) a security relevant configuration. Furthermore, since the 
$\bot$ value is an artifical ``error'' value of the security automaton with
no Java counterpart, we know that if $\MS = \MS^g$, then $\MS^g \neq \bot$. 
Thus, by lemma \ref{lem:ghost_mon_correct}, $\SRT(E) \in \Contract$ and 
therefore $P$ adheres to $\Contract$.
\end{proof}

\subsection{Example Proof Generation}
The process of generating contract adherence proofs is closely 
related to the process of embedding the reference monitor, thus the 
inlining and proof generation is preferrably done by the same agent. 
This section describes how proofs may be generated for code produced by 
the example inliner presented in Section \ref{sect:inlining}.

The monitor invariant, $\MS = \MS^g$ is set as each methods pre- and 
post-condition. The assertion for each specific instruction is generated 
differently, according to whether the instruction appears as part of an 
inlined block or not. Instructions inside the inlined block affect the 
processing of the embedded state, method call arguments etc. For this 
reason these instructions need detailed analysis using the $\WP$ function. 
Instructions outside the inlined blocks, on the other hand, allow a more 
robust treatment, as they are only required to preserve the monitor invariant 
which they do (see fact \ref{fact:putstatic} in Appendix \ref{app:example_inliner}). The critical property of 
the annotation function is the following:
\begin{lemma}\label{lem:ex_proof_gen}
Given a method $M = (I, H)$ of ${\cal I}^g(P, \Contract)$ and a set $\IL$ 
labelling the inlined instructions in $I$, an array $A$ of assertions can 
be computed such that the extended method $(I,H,A, \MS=\MS^g, \MS=\MS^g)$ 
is locally valid.
\end{lemma}
\begin{proof}
A general construction is illustrated in Appendix \ref{app:proof_gen_proof}.
\end{proof}
The array is constructed by annotating the return 
instructions with the post-condition, and then in a breadth first manner, 
annotate the preceeding instructions using the $\WP$ function in case of 
inlined instructions and by using the monitor invariant in other cases.

\begin{theorem}[Proof Generation]\label{thm:proof_generation}
For each program $P$ and contract $\Contract$ there is an algorithm, 
polynomial in $\Size{P}+\Size{\Contract}$, which produces an adherence proof 
of $\Inline(P, \Contract)$.
\end{theorem}
\begin{proof}
The algorithm described above treats each method in isolation. The 
breadth first traversal of the instructions takes time linearly 
proportional to the size of the instruction array plus the number of 
ghost updates. The resulting adherence proof is correct by construction.
\end{proof}
As an example Figure \ref{fig:mod_open_data} illustrates a generated proof 
for a part of a program which has been inlined to comply with the 
policy in Figure \ref{fig:demo_secspec}.
\begin{figure}[!ht]
\centering
{\ttfamily
\begin{tabular}{@{~~}l@{~~}}
SCOPE Session\\
\\
SECURITY STATE boolean haveRead = false;\\
\\
BEFORE javax.microedition.rms.RecordStore\\
\phantom{~~~~~~~~}.openRecordStore(string name,\\
\phantom{~~~~~~~~}boolean createIfNecessary)\\
\phantom{~~~~}PERFORM\\
\phantom{~~~~~~~~}true -> \verb+{+ haveRead = true; \verb+}+\\
\\
BEFORE javax.microedition.io.Connector\\
\phantom{~~~~~~~~}.openDataOutputStream(string url) \\
\phantom{~~~~}PERFORM\\
\phantom{~~~~~~~~}haveRead == false -> \verb+{ }+\\
\end{tabular}
}
\caption{\label{fig:demo_secspec} A ConSpec specification which disallows the program from sending data over the network after accessing phone memory.}
\end{figure}

\begin{figure}[!ht]
\centering
{\small\ttfamily\begin{tabular}{@{}l@{}r@{}l@{}}
  &          & $\vdots$                                                       \\
  &      40: & $\{ \Psi \}$                                                   \\
  &          & aload\textunderscore 1                                         \\
\multirow{22}{*}{\rmfamily inlined $\left\{ \begin{tabular}{@{}l@{}} \\ \\ \\ \\ \\ \\ \\ \\ \\ \\ \\ \\ \\ \\ \\ \\ \\ \\ \\ \\ \\ \\ \end{tabular} \right.$}
  &      41: & $\{ \IF(0 \neq {\tt SS.haveRead}, \True,$                  \\
  &          & $\phantom{\{} \IF(\ghost{haveRead} = \False, \Psi, \bot = {\tt SS.haveRead})) \}$ \\\
  &          & astore\textunderscore 3                                        \\
  &      42: & $\{ \IF(0 \neq {\tt SS.haveRead}, \True, $ \\\
  &          & $\phantom{\{} \IF(\ghost{haveRead} = \False, \Psi, \bot = {\tt SS.haveRead})) \} $ \\
  &          & getstatic SS.haveRead                                    \\
  &      45: & $\{ \IF(0 \neq \Stack_0, \True, $ \\\
  &          & $\phantom{\{} \IF(\ghost{haveRead} = \False, \Psi, \bot = {\tt SS.haveRead})) \} $ \\
  &          & iconst\textunderscore 0                                        \\
  &      46: & $\{ \IF(\Stack_0 \neq \Stack_1, \True, $ \\
  &          & $\phantom{\{} \IF(\ghost{haveRead} = \False, \Psi, \bot = {\tt SS.haveRead})) \}$ \\
  &          & if\textunderscore icmpne 52                                    \\
  &      49: & $\{ \IF(\ghost{haveRead} = \False, \Psi, \bot = {\tt SS.haveRead}) \} $ \\
  &          & goto 56                                                        \\
  &      52: & $\{ \True \}$                                                  \\
  &          & iconst\textunderscore m1                                       \\
  &      55: & $\{ \True \}$                                                  \\
  &          & invokestatic System.exit                                       \\
  &      56: & $\{ \IF(\ghost{haveRead} = \False, \Psi, \bot = {\tt SS.haveRead}) \}$ \\
  &          & aload\textunderscore 3                                         \\
  &          & $ \{ \IF(\ghost{haveRead} = \False, \Psi, \bot = {\tt SS.haveRead}) \}$ \\
  &          & $ \langle \ghost{haveRead} := \ghost{haveRead} = \False \rightarrow \ghost{haveRead} \rangle $ \\
  &      71: & $\{ \Psi \}$                                                   \\
  &          & invokestatic                                                   \\
  &          & ~~~~Connector.openDataOutputStream                             \\
  &      74: & $\{ \Psi \}$                                                   \\
  &          & astore\textunderscore 2                                        \\
  &          & $\vdots$                                                       \\
\end{tabular}}
\caption{\label{fig:mod_open_data} Generated assertions for inlining of {\tt Connect\-or.openDataOutputStream} where $\Psi$ denotes the monitor invariant.}
\end{figure}

\subsection{Proof Recognition}
Checking the validity of contract adherence proofs involves verifying 
local validity, which in general is undecidable. However, the problem 
is much simplified in our setup, since proofs apply to programs that 
have already been inlined. Validity checking may still be hard or impossible, 
however, due to the use of primitive data types with difficult equational 
theories. For this reason the theorem below is restricted to contracts over
freely generated theories.
%
%
%
\begin{theorem}[Efficient Recognition]\label{thm:eff_recog}
The class of adherence proofs generated from programs inlined with contracts
over a freely generated theory is recognizable in polynomial time.
\end{theorem}
%
\begin{proof}
To verify the validity of a given adherence proof we look at the requirements of definition \ref{def:proof_validity}. Verifying that the pre- and post-conditions equal the monitor invariant is a simple syntactic check and can be done in time linearly proportional to the number of methods in $P$.

For the requirement of local validity, it is sufficient to check that the verification conditions 1 and 2 from definition \ref{def:local_validity} can be rewritten to $\True$ in time polynomial in the size of the instruction array. The interesting verification conditions are those of the form $A_L \Rightarrow \WP_M(L)$ where $L$ is the label of the first instruction in an inlined block. $A_L$ is, in this case, of the form $ \MS = \MS^g \AND \ghost{a}_0 = a_0 = s_0 \AND \ldots \AND \ghost{a}_m = a_m = s_m $ and $\WP_M(L)$ is of the form 

\begin{tabbing}
xx\=xx\=xx\=xx\= \kill
$\SELECT((t:c^n \AND \ghost{t}:c^n, \ldots, t:c^1 \AND \ghost{t}:c^1),$\\
          \>$(\SELECT((c^n.m_{G_1} \AND c^n.m\ghost{}_{G_1},
                             \ldots,c^n.m_{G_i}\AND c^n.m\ghost{}_{G_i}), $\\
          \>   \>          \>$(c^n.m_{f_1}(\MS, \mathbf{a})
             =c^n.m\ghost{}_{f_1}(\ghost{\MS}, \ghost{\mathbf{a}}),\ldots,$\\
          \>   \>          \>         \> $c^n.m_{f_i}(\MS, 
                      \mathbf{a})=c^n.m\ghost{}_{f_i}(\ghost{\MS}, 
                                             \ghost{\mathbf{a}})), \True),$\\
          \>   \> ~~~$\vdots$ \>      \> ~~~$\vdots$\\
          \>   \>$\SELECT((c^1.m_{G_1} \AND 
                      c^1.m\ghost{}_{G_1},\ldots,c^1.m_{G_j}\AND 
                                                c^1.m\ghost{}_{G_j}), $\\
          \>   \>          \>         \> $c^1.m_{f_1}(\MS, \mathbf{a})=
                               c^1.m\ghost{}_{f_1}(\ghost{\MS}, 
                                        \ghost{\mathbf{a}}),\ldots,$\\
          \>   \>          \>         \> $c^1.m_{f_j}(\MS, \mathbf{a})=
                           c^1.m\ghost{}_{f_j}(\ghost{\MS}, 
                                            \ghost{\mathbf{a}})), \True)), $\\
          \>$\MS = \ghost{\MS})$
\end{tabbing}

The verification condition can then be rewritten and simplified by 
iterated applications of the rule $x = y \Rightarrow \phi \longrightarrow \phi[z/x][z/y]$ where $x$ and $y$ are instantiated with real variables and ghost counterparts respectively and where $z$ does not occur in $\phi$. 
These rewrites can be performed in time proportional to the
length of the formula and does not increase the 
size of the expression since $x$, $y$ and $z$ are atomic. The result can then be rewritten to $\True$ using the rules $(\psi \Rightarrow \phi) \AND (\neg\psi \Rightarrow \phi) \longrightarrow \phi$ and $\phi = \phi \longrightarrow \True$ in time polynomial in the size of the formula.

All other verification conditions ($\Pre_M \Rightarrow A_0$, $A_L \Rightarrow \WP_M(L)$ for all labels $L$ except those of the first instructions in an inlined block are trivial as their antecedents and succeedents are identical.
\end{proof}

\section{Implementation and Evaluation}
A full implementation of the framework, including a Java SE proof generator, a Java ME client, instructions and examples is available at \url{www.csc.kth.se/~landreas/irm_pcc}. Both the on- and the off-device software utilize a parser generated by CUP / JFlex~\cite{cup,jflex} and the ASM library~\cite{asm} for handling class files. Table \ref{tab:benchmarks} summarizes overhead for inlining, proof generation and load-time proof recognition on two example applications and policies:
\begin{itemize}
\item \emph{MobileJam}: A GPS based traffic jam reporter which utilizes the Yahoo! Maps API.\\Policy: Only connect to \url{http://local.yahooapis.com}.
\item \emph{Snake}: A classic game of snake in which the player may submit current score to a server.\\Policy: Do not send data over network after reading from phone memory.
\end{itemize}
\begin{table}
\center{\begin{tabularx}{\linewidth}{Xr@{~}lr@{~}l} \hline
                                 & \multicolumn{2}{l}{MobileJam}    & \multicolumn{2}{l}{Snake} \\ \hline
Security Relevant Invokes        &     4 &      &    2 &          \\
Original Size                    & 428.0 & kb   & 43.7 & kb       \\
Size increase for IRM            &   4.8 & kb   &  1.1 & kb       \\
Size increase for Proofs         &  20.6 & kb   &  2.6 & kb       \\
Inlining                         &  10.1 & s    &  8.6 & s        \\
Proof Generation                 &   4.7 & s    &  0.8 & s        \\
Proof Recognition                &    98 & ms   &  117 & ms       \\
\end{tabularx}}
\caption{\label{tab:benchmarks} Benchmarks for the two case studies.}
\end{table}

Inlining and proof generation was performed on an Intel Core 2 CPU at 1.83~GHz with 2~Gb memory and proof recognition was performed on a Sony-Ericsson W810i. The implementation is to be considered a prototype, and very few optimizations in terms of e.g. proof size have been implemented.

\section{Conclusions}
We have demonstrated the feasibility of a proof-carrying approach to certified monitor inlining at the level of practical Java bytecode, including exceptions and inheritance. This answers partially a question raised by K.~W. Hamlen et al.~\cite{HamMor06}.

We have proved correctness of our approach in the sense of soundness: Contract adherence proofs are sufficient to ensure compliance. This soundness proof has been formalized \cite{coqscript} in Coq.
We also obtain partial completeness results, namely that proofs for inlined 
programs can always be generated, and such proofs are guaranteed to be 
recognized at program loading time, at least when contracts do not use
equational tests that are too difficult. Other properties are also interesting 
such as transparency \cite{saltzer75}, roughly, that all adherent behaviour is preserved by the inliner. This type of property is, however, more relevant for the specific inliner, and not so much for the certification mechanism, and consequently not addressed here (but see e.g. \cite{Lig06,VanPie08,DamJLP10,DamJLP09} for results in  this direction).

The approach is efficient: Proofs are small and recognised easily, by a simple proof checker. An interesting feature of our approach is that detailed modelling of bytecode instructions is needed only for instructions appearing in the inlined code snippets. For other instructions a simple conditional invariance property on static fields of final objects suffices. This means, in particular, that our approach adapts to new versions of the Java virtual machine very easily, needing only a check that the static field invariance is maintained. Worth pointing out also is that the enforcement architecture can be realized in a way which is backwards compatible, in the sense that PCC-aware client programs can be executed without modification in a PCC-unaware host environment.

It is possible to extend our framework to multi-threading by protecting security relevant updates with locks, either locking the entire inlined block or releasing the lock during the security relevant call itself for increased parallellism. For proof generation the main upshot is that assertions must be stable under interference by other threads. Briefly, this requires the ability to protect fields, such as those in the security state class, with locks by only allowing updates of these fields when the lock has been acquired. The validity of an assertion may then only depend on fields protected by locks that has been acquired at that point in the code. This work is currently in progress.



\bibliography{text}

\clearpage
\appendix

\section{Implementation of the Example Inliner}\label{app:example_inliner}
Our inliner lets the state of the embedded security monitor be represented by a static field $\MS$ of a final security state class, named to avoid clashes with classes in the target program. This choice of representation relies on the following fact of JVM execution and allows for our open-ended treatment of large parts of the instruction set.
\begin{fact}\label{fact:putstatic}
Suppose $c$ is final and $f$ is static. If $C=(h, (M, \PC,$ $s,r)::R)\transJVM C'$ and $M[\PC] \neq {\tt putstatic}\ c.f$, then $\Sem{c.f}C = \Sem{c.f}C'$.
\end{fact}
In other words, the only instruction which can affect the value stored in a static field $f$ of a final class $c$ is an explicit assignment to $c.f$. In particular, the assumption ensures that instructions originating from the target program are unable to affect the embedded monitor state.

For simplicity we assume (without loss of generality) that ConSpec policies initialize the security state variables to the default Java values.

Each {\tt invokevirtual} $c.m$ instruction is replaced by a block of inlined code that evaluates which concrete method is being invoked, then checks and updates the security state accordingly. We assume for simplicity that no instructions in a block of inlined code other than {\tt athrow} will raise exceptions. The code is easily adapted at the cost of some additional complexity to take runtime exceptions violating this assumption into account. 

Figure \ref{fig:schematic_conspec} shows a schematic policy for a method $m : {\tt int} \rightarrow {\tt int}$ defined in class a $c$ and overridden in a subclass $d$. The policy has event clauses for {\sc before}, {\sc after} and {\sc exceptional} cases for each definition of $m$, each with two guards and two statement lists. 
\begin{figure}[h]
\centering
{\scriptsize\ttfamily
\begin{tabular}{@{}l@{}}
SCOPE Session\\
SECURITY STATE int ms = 0; \\[2mm]
BEFORE~~~~~ c.m(int a)\
PERFORM cb$_{g_1}$ -> \verb+{+cb$_{s_1}$\verb+}+ | cb$_{g_2}$  ->  \verb+{+cb$_{s_2}$\verb+}+ \\
AFTER~~~r = c.m(int a)\
PERFORM ca$_{g_1}$ -> \verb+{+ca$_{s_1}$\verb+}+ | ca$_{g_2}$  ->  \verb+{+ca$_{s_2}$\verb+}+ \\
EXCEPTIONAL c.m(int a)\
PERFORM ce$_{g_1}$ -> \verb+{+ce$_{s_1}$\verb+}+ | ce$_{g_2}$  ->  \verb+{+ce$_{s_2}$\verb+}+ \\[2mm]
BEFORE~~~~~ d.m(int a)\
PERFORM db$_{g_1}$ -> \verb+{+db$_{s_1}$\verb+}+ | db$_{g_2}$  ->  \verb+{+db$_{s_2}$\verb+}+ \\
AFTER~~~r = d.m(int a)\
PERFORM da$_{g_1}$ -> \verb+{+da$_{s_1}$\verb+}+ | da$_{g_2}$  ->  \verb+{+da$_{s_2}$\verb+}+ \\
EXCEPTIONAL d.m(int a)\
PERFORM de$_{g_1}$ -> \verb+{+de$_{s_1}$\verb+}+ | de$_{g_2}$  ->  \verb+{+de$_{s_2}$\verb+}+
\end{tabular}}
\caption{\label{fig:schematic_conspec} Schematic ConSpec policy}
\end{figure}

Figure \ref{fig:inlining_schema} gives the inlining details for the policy schema in Figure \ref{fig:schematic_conspec}. In the figure, each {\slshape \ttfamily [EVALUATE $g$]} section transforms a configuration $(h, (M, pc, s, r)::R)$ to $(h, (M, pc', v$ $::s, r)::R)$ where $v$ is 0 or 1 if the guard $g$ is false or true respectively. An {\slshape \ttfamily [EXECUTE ${\it stmts}$]} transforms the configuration $(h, (M, pc, s, r)::R)$ to $(h[\llbracket stmts \rrbracket (\MS) /\MS],(M,$ $pc', s, r)::R)$.

The remaining invoke instructions ({\tt invokestatic}, {\tt in\-vokeinterface} and {\tt invokespecial}) can be handled similarly.

\begin{figure}[!ht]
\centering
{\footnotesize\ttfamily
\begin{tabular}{@{}r @{~} l @{$\>$}|@{$\>\>$} r @{~} l@{}}
tArgs:        & astore $r_a$                             & deFail:       & iconst\textunderscore 1                   \\
              & astore $r_t$                             &               & inv\_static Sys.exit                      \\
              & aload $r_t$                              & ceChk:        & aload $r_t$                               \\
              & aload $r_a$                              &               & instanceof c                              \\
dbChk:        & aload $r_t$                              &               & ifeq EEnd                                 \\
              & instanceof d                             & ceGrd1:       & {\slshape [EVALUATE ce$_{g_1}$]}          \\
              & ifeq cbChk                               &               & ifeq ceGrd2                               \\
dbGrd1:       & {\slshape [EVALUATE db$_{g_1}$]}         &               & {\slshape [EXECUTE ce$_{s_1}$]}           \\
              & ifeq dbGrd2                              &               & goto EEnd                                 \\
              & {\slshape [EXECUTE db$_{s_1}$]}          & ceGrd2:       & {\slshape [EVALUATE ce$_{g_2}$]}          \\
              & goto BEnd                                &               & ifeq ceFail                               \\
dbGrd2:       & {\slshape [EVALUATE db$_{g_2}$]}         &               & {\slshape [EXECUTE ce$_{s_2}$]}           \\
              & ifeq dBFail                              &               & goto EEnd                                 \\
              & {\slshape [EXECUTE db$_{s_2}$]}          & ceFail:       & iconst\textunderscore 1                   \\
              & goto BEnd                                &               & inv\_static Sys.exit                      \\
dBFail:       & iconst\textunderscore 1                  & EEnd:         & athrow                                    \\
              & inv\_static Sys.exit                     & hdlEnd:       & aload $r_t$                               \\
cbChk:        & aload $r_t$                              &               & instanceof d                              \\
              & instanceof c                             &               & ifeq caChk                                \\
              & ifeq BEnd                                & daGrd1:       & {\slshape [EVALUATE da$_{g_1}$]}          \\
cbGrd1:       & {\slshape [EVALUATE cb$_{g_1}$]}         &               & ifeq daGrd2                               \\
              & ifeq cbGrd2                              &               & {\slshape [EXECUTE da$_{s_1}$]}           \\
              & {\slshape [EXECUTE cb$_{s_1}$]}          &               & goto AEnd                                 \\
              & goto BEnd                                & daGrd2:       & {\slshape [EVALUATE da$_{g_2}$]}          \\
cbGrd2:       & {\slshape [EVALUATE cb$_{g_2}$]}         &               & ifeq daFail                               \\
              & ifeq cbFail                              &               & {\slshape [EXECUTE da$_{s_2}$]}           \\
              & {\slshape [EXECUTE cb$_{s_2}$]}          &               & goto AEnd                                 \\
              & goto BEnd                                & daFail:       & iconst\textunderscore 1                   \\
cbFail:       & iconst\textunderscore 1                  &               & inv\_static Sys.exit                      \\
              & inv\_static Sys.exit                     & caChk:        & aload $r_t$                               \\
BEnd:         & invokevirtual c.m                        &               & instanceof c                              \\
              & goto hdlEnd                              &               & ifeq AEnd                                 \\
hdlStrt:      & aload $r_t$                              & caGrd1:       & {\slshape [EVALUATE ca$_{g_1}$]}          \\
              & instanceof d                             &               & ifeq caGrd2                               \\
              & ifeq ceChk                               &               & {\slshape [EXECUTE ca$_{s_1}$]}           \\
deGrd1:       & {\slshape [EVALUATE de$_{g_1}$]}         &               & goto AEnd                                 \\
              & ifeq deGrd2                              & caGrd2:       & {\slshape [EVALUATE ca$_{g_2}$]}          \\
              & {\slshape [EXECUTE de$_{s_1}$]}          &               & ifeq caFail                               \\
              & goto EEnd                                &               & {\slshape [EXECUTE ca$_{s_2}$]}           \\
deGrd2:       & {\slshape [EVALUATE de$_{g_2}$]}         &               & goto AEnd                                 \\
              & ifeq deFail                              & caFail:       & iconst\textunderscore 1                   \\
              & {\slshape [EXECUTE de$_{s_2}$]}          &               & inv\_static Sys.exit                      \\
              & goto EEnd                                & AEnd:         &                                           \\
\end{tabular}}
\caption{\label{fig:inlining_schema} Schematic inlining of policy in Figure \ref{fig:schematic_conspec}}
\end{figure}

\clearpage
\section{Proof of Lemma \ref{lem:ex_proof_gen}}\label{app:proof_gen_proof}
Figure \ref{fig:schematic_annotations} shows the construction for a call of a method  $m : {\tt int} \rightarrow {\tt int}$ in class $c$, under the schematic contract shown in Figure \ref{fig:schematic_contract}. We assume that an exception thrown by the invoked method is matched by an exception handler table entry on the form $(30, 32, 34, \mathit{any})$. For brevity we let $\sigma_{\it bef}$, $\sigma_{\it aft}$ and $\sigma_{\it exc}$ denote the appropriate substitution for the effect of updating $\MS$ according to the before, after and exceptional clause of $c.m$ respectively. For instance, if {\tt bef$_s$} denotes {\tt ms = ms * x; ms = ms - 5}, then $\sigma_{\it bef}$ is $[(\MS \cdot x) - 5 / \MS]$.
\begin{figure}[!ht]
\centering
{\ttfamily
\begin{tabular}{@{}l@{}}
SCOPE Session\\
SECURITY STATE DECLARATION\\[2mm]

BEFORE~~~~~ c.m(int a) PERFORM bef$_g$ -> \verb+{+bef$_s$\verb+}+ \\
AFTER~~~r = c.m(int a) PERFORM aft$_g$ -> \verb+{+aft$_s$\verb+}+ \\
EXCEPTIONAL c.m(int a) PERFORM exc$_g$ -> \verb+{+exc$_s$\verb+}+
\end{tabular}
}
\caption{\label{fig:schematic_contract} Schema contract for the proof of Lemma \ref{lem:ex_proof_gen}.}
\end{figure}
\begin{figure}
{\scriptsize
\newlength{\sep}
\setlength{\sep}{0.6mm}
\ttfamily
\begin{center}
\begin{tabular}{@{}l@{~}l@{}}

        & $\MS = \MS^g$ \\
        & NON-INLINED INSTRUCTION \\[\sep]
        & // INLINED CODE START \\[\sep]
        
        & $\MS = \MS^g$ \\
        & ASTORE a\\
        & ASTORE t\\
        & ALOAD t \\
        & ALOAD a \\[\sep]
        
        & // BEFORE \\[\sep]

26:     & $\IF(t:c, A_{28}, A_{30})$ \\
        & ALOAD t \\
        & INSTANCEOF c \\
        & IFEQ 30 \\[\sep]
        
28:     & $\IF(\mbox{bef}_g, \IF(s_1:c, \IF(\mbox{bef}_g, \MS\sigma_{\it bef}(a) = \MS^g\sigma_{\it bef}(s_0), \MS\sigma_{\it bef} = \bot), $\\
        & $\MS = \MS^g) \AND a = s_0 \AND t = s_1, \True) $ \\
        & [EVALUATE bef$_g$] \\
        & IFEQ 29 \\
        & [PERFORM bef$_s$] \\
        & GOTO 30 \\[\sep]
        
29:     & $\True$ \\
        & ICONST\_1 \\
        & INVOKESTATIC System.exit \\[\sep]

30:     & $\IF(s_1:c, \IF(\mbox{bef}_g, \MS = \MS^g\sigma_{\it bef}(s_0), \MS = \bot), \MS = \MS^g)\ \AND $\\
        & \hphantom{~~~~} $a = s_0 \AND t = s_1 $ \\
        & $\langle (t^g, a^g) := (s_1, s_0) \rangle$ \\
        & $\langle \MS^g := t^g : c \rightarrow \delta(\MS^g, (c.m, a^g)^\uparrow) \mid \MS^g \rangle$ \\[\sep]
        
        & $\MS = \MS^g \AND a = a^g \AND t = t^g $ \\
        & INVOKEVIRTUAL c.m(int) : int \\[\sep]
        
32:     & $\MS = \MS^g \AND a = a^g \AND t = t^g $ \\
        & $ \langle r^g := s_0 \rangle$ \\
        & $ \langle \MS^g := t^g : c \rightarrow \delta(\MS^g, (c.m, a^g, r^g)^\downarrow) \mid \MS^g \rangle$ \\[\sep]
        
        & $A_{43}[r/\Stack_0]$ \\
        & ASTORE r \\
        & ALOAD r \\[\sep]
        
        & $A_{43}$ \\
        & GOTO 43 \\[\sep]

        & // EXCEPTIONAL \\[\sep]

34:     & $\MS = \MS^g \AND a = a^g \AND t = t^g $ \\
        & $ \langle \MS^g := t^g:c \rightarrow \delta(\MS^g, (c.m, a^g)^\Downarrow) \mid \MS^g \rangle$ \\[\sep]
        
38:     & $\IF(t : c, A_{40}, A_{42})$ \\
        & ALOAD t \\
        & INSTANCEOF c \\
        & IFEQ 42 \\[\sep]

40:     & $\IF(\mbox{exc}_g, \MS\sigma_{\it exc}(a) = \MS^g, A_{41})$ \\
        & [EVALUATE exc$_g$] \\
        & IFEQ 41 \\
        & [PERFORM exc$_s$] \\
        & GOTO 42 \\[\sep]

41:     & $\True$ \\
        & ICONST\_1 \\
        & INVOKESTATIC System.exit \\[\sep]

42:     & $\MS = \MS^g$ \\
        & ATHROW \\[\sep]

        & // AFTER \\[\sep]

43:     & $\IF(t:c, A_{44}, A_{46})$ \\
        & ALOAD t \\
        & INSTANCEOF c \\
        & IFEQ 46 \\[\sep]

44:     & $\IF(\mbox{aft}_g, \MS\sigma_{\it aft}(r, a) = \MS^g, \True)$ \\
        & [EVALUATE aft$_g$] \\
        & IFEQ 45 \\
        & [PERFORM aft$_s$] \\
        & GOTO 46 \\[\sep]
        
45:     & $\True$ \\
        & ICONST\_1 \\
        & INVOKESTATIC System.exit\\[\sep]
        
        & // INLINING END \\[\sep]
        
46:     & $\MS = \MS^g$ \\
        & NON-INLINED INSTRUCTION
\end{tabular}
\end{center}
}
\caption{\label{fig:schematic_annotations} Schematic annotation for contract displayed Figure \ref{fig:schematic_contract}}
\end{figure}


%
%

\end{document}